\documentclass[aps,pra,twocolumn,amsmath,showpacs]{revtex4}
\usepackage{bm}
\usepackage{graphicx}
\begin{document}
\setlength{\arraycolsep}{2pt}
\title{Distinguishing two single-mode Gaussian states by homodyne detection:\\ 
An information-theoretic approach}
\author{Hyunchul Nha and H. J. Carmichael}
\affiliation{Department of Physics, University of Auckland, Private Bag 92019,
Auckland, New Zealand} 

\begin{abstract}
It is known that quantum fidelity, as a measure of the closeness of two quantum
states, is operationally equivalent to the minimal overlap of the probability
distributions of the two states over all possible POVMs; the POVM realizing the
minimum is optimal. We consider the ability of homodyne detection to distinguish
two single-mode Gaussian states, and investigate to what extent it is optimal in
this information-theoretic sense. We completely identify the conditions under which
homodyne detection makes an optimal distinction between two single-mode Gaussian
states of the same mean, and show that if the Gaussian states are pure,
they are always optimally distinguished.
\end{abstract}
\maketitle

\narrowtext
\section{Introduction}
In the field of information processing, both classical and quantum, it is crucial to
be able to distinguish between two items of information. Various distance measures have been
proposed to quantify the separation of items of information, and in particular, the
trace distance and quantum fidelity are widely used in the quantum information field
\cite{qip}. The quantum fidelity was first defined by Jozsa \cite{Jozsa}, based on
Uhlmann's transition probability \cite{Uhlmann}: given two quantum states, $\rho_1$
and $\rho_2$, it is given by $F(\rho_1,\rho_2)={\rm tr}
\sqrt{\mkern-3mu\sqrt{\rho_1}\mkern3mu\rho_2\sqrt{\rho_1}}$. The fidelity quantifies
the closeness of the two states, so that their separation can be measured, e.g.,
by the Bures distance $D_B^2=2(1-F)$~\cite{Bures}. 

The quantum fidelity can be given a physical or operational significance as follows
\cite{Fuchs}. By performing a general POVM measurement $\{E_m\}$, where the $E_m$
sum to the identity operator \cite{Peres}, one obtains probability distributions
$p_1(m)\equiv{\rm tr}(\rho_1E_m)$ and $p_2(m)\equiv{\rm tr}(\rho_2E_m)$ for the
states $\rho_1$ and $\rho_2$, respectively. It is known that the overlap of these
distributions is always greater than or equal to the fidelity---i.e.,
$\sum_m\sqrt{p_1(m)}\sqrt{p_2(m)}\ge F$. In particular, Fuchs and Caves have
proved that there always exists a POVM satisfying the equality \cite{Fuchs,Barnum}. 
The fidelity, then, is equal to the minimal overlap of the defined probability 
distributions over all possible POVMs, $F={\rm min}_{\{E_m\}}\sum_m\sqrt{p_1(m)}
\sqrt{p_2(m)}$. 

In the proof given by Fuchs and Caves \cite{Fuchs}, the optimal POVM---that
yielding the quantum fidelity---depends on the pair of states being compared
\cite{Fuchs}. In experiments, however, one would prefer to take a predetermined
class of measurements, rather than set up the measurement case by case. 
Then, the pairs of quantum states that can be optimally distinguished 
by the chosen measurements will be restricted in general, 
and the condition for such pairs depends on the kind of measurement under consideration. 
In this work, we consider homodyne detection, a continuous variable (CV) measurement,
as a means of distinguishing two single-mode Gaussian states in the aforementioned
information-theoretic sense. There has been significant progress in quantum information
processing using continuous variable systems \cite{QIbook}, and
much attention has been paid to their Gaussian states which are rather easily
accessible in experiments and tractable in theoretical calculations.

In principle, optical tomography by homodyne detection can characterize the state
of an optical field completely \cite{Leonhardt}. Working in this direction,
Kim {\it et al.} \cite{Kim} have proposed an experiment to obtain the fidelity
directly by mixing the two fields to be distinguished at a beam splitter and
measuring the Wigner function of the output. The method is valid so long as
one of the field states is pure. Our proposal, instead, is to perform
independent homodyne measurements on the two fields, and determine 
their closeness from the probability distributions of the measurement results. 
Homodyne detection is highly efficient compared with photon counting, and more
importantly, it is genuinely a CV measurement, in the sense
that it reveals the continuous (wave) nature of the field. We aim to determine
to what extent such a measurement is optimal in distinguishing Gaussian states.
In general, homodyne detection refers to any measurement schemes 
where the signal optical field is superposed with an auxiliary mode 
that has the same central frequency as the signal field. 
In this paper, we consider the usual balanced homodyne detection (BHD) where 
a very strong local oscillator field is mixed with the signal. 
The quadrature amplitudes $\hat X_{\phi}\equiv
(\hat ae^{-i\phi}+\hat a^{\dag}e^{i\phi})/2$ are then measured in this BHD, 
where $\hat a$ is the annihilation operator for the optical field and 
$\phi$ is an adjustable phase.
We compare the quantum fidelity with the minimum overlap $I_{\phi}\equiv
\int dx_\phi \sqrt{p_1(x_\phi)}\sqrt{p_2(x_\phi)}$, the minimum taken with
respect to all measurement angles~$\phi$.

Twamley has calculated the fidelity of two undisplaced thermal states
\cite{Twamley}, and Paraoanu and Scutaru that of displaced thermal states
\cite{Paraoanu}. Scutaru has obtained the fidelity of arbitrary
single-mode Gaussian states \cite{Scutaru}. We consider the set of all pairs of
single-mode Gaussian states that have identical values for the means of the
quadrature amplitudes, i.e., $\langle\hat X_\phi\rangle_{\rho_1}=\langle
\hat X_\phi\rangle_{\rho_2}$. We identify the conditions under which homodyne
detection makes the optimal distinction between these states. While the
considered set of states is restricted, we note that many CV quantum
information protocols do not change the mean amplitude of the field.
Examples include (imperfect) quantum teleportation \cite{Braunstein} and
Gaussian cloning \cite{Cerf}, each with unity gain. 

The paper is structured as follows. In Sec.~II we briefly review the basic
formalism of Gaussian states and the calculation of the fidelity for a
general pair of Gaussian states. Homodyne detection is considered in Sec.~III,
where we construct the overlap of the probability distributions for Gaussian
states of equal mean. The minimal overlap is compared with the quantum fidelity 
to identify those cases in which homodyne detection is optimal in distinguishing
the states. Pairs of radially symmetric (unsqueezed) Gaussian states with
different means are also briefly considered. A summary and discussion is
presented in Sec.~IV. 
 
\section{Fidelity of two single-mode Gaussian states}
\subsection{Gaussian states}
A Gaussian state $\rho$ is completely characterized by its first and the second
moments. It possesses a Gaussian characteristic function,
\begin{equation}
C({\bm x})\equiv{\rm tr}\{\rho\hat W({\bm x})\}=e^{i\langle\hat{\bm R}\rangle
\mkern1mu{\bm x}^T}e^{-\frac{1}{4}{\bm x}{\bm \Gamma}{\bm x}^T},
\label{eqn:characteristic}
\end{equation}
where $\hat{\bm R}\equiv\left({\hat q},{\hat p}\right)$, with
$\hat q$ and $\hat p$ position and momentum operators, respectively,
${\bm x}$ is a real 2-dim row vector, and $\hat W({\bm x})=e^{i\hat{\bm R}\cdot
{\bm x}^T}$ is the Weyl operator. The canonical operators ${\hat q}$ and ${\hat p}$
are related to the quadrature amplitudes $\hat X\equiv(\hat a+\hat a^{\dag})/2=
\hat q/\sqrt2$ and $\hat Y\equiv-i(\hat a-\hat a^{\dag})/2=\hat p/\sqrt2$. The
covariance matrix ${\bm \Gamma}$ is real, symmetric, and positive-definite, with
matrix elements
\begin{equation}
\Gamma_{ij}=\langle\Delta\hat R_i\Delta\hat R_j+\Delta\hat R_j\Delta\hat R_i\rangle,
\quad (i,j)=1,2.
\end{equation}
The Heisenberg uncertainty relation must be incorporated in order for ${\bm \Gamma}$ 
to represent a legitimate physical state. Thus, with the canonical commutation relations
represented by the symplectic matrix ${\bm \Sigma}$, with $[\hat R_i,\hat R_j]=
i\Sigma_{ij}$, i.e.,
\begin{equation}
{\bm\Sigma}=
\begin{pmatrix}
{0}&{1}\cr{-1}&{0}
\end{pmatrix},
\end{equation}
the matrix ${\bm\Gamma}+i{\bm \Sigma}$ ($\Gamma_{ij}+i\Sigma_{ij}=2\langle
\Delta\hat R_i\Delta\hat R_j\rangle$) is required to be positive semidefinite.
The condition is necessary and sufficient for the covariance matrix ${\bm \Gamma}$
to represent a physical state~\cite{Simon1}.

Consider now a real linear transformation ${\bm S}$ from the canonical operators
$\hat {\bm R}$ to another set of operators ${\hat{\bm R}}^\prime$, with
$\hat{\bm R}^{\prime T}={\bm S}\hat{\bm R}^T$. If the canonical commutation relations are
to be preserved ($[\hat R_i',\hat R_j']=i\Sigma_{ij}$), ${\bm S}$ must satisfy
the relation ${\bm S}{\bm\Sigma}{\bm S}^T={\bm\Sigma}$. Such so-called symplectic
transformations correspond to unitary Gaussian operations generated by Hamiltonians
quadratic in the operators $\hat{\bm R}$ \cite{Simon1}. The transformation ${\bm S}$
maps a real covariance matrix ${\bm\Gamma}$ to another such matrix, 
\begin{eqnarray}
{\bm\Gamma}\rightarrow{\bm\Gamma}'={\bm S}{\bm\Gamma}{\bm S}^T.
\label{eqn:cov_trans}
\end{eqnarray} 
If ${\bm\Gamma}+i{\bm\Sigma}\ge{\bm 0}$, then 
${\bm\Gamma}'+i{\bm\Sigma}={\bm S}({\bm\Gamma}+i{\bm \Sigma}){\bm S}^T\ge{\bm 0}$; 
a physical covariance matrix remains physical under a symplectic transformation.

Since five independent real parameters fully determine the characteristic
function---three for the covariance matrix $\Gamma$ and two for the mean values
$\langle\hat{\bm R}\rangle$---it is not difficult to see that a general single-mode
Gaussian state can be parametrically represented by a squeezed and displaced thermal
state \cite{Marian}. We may write the density operator in the form 
\begin{eqnarray}
\rho=\hat D(\alpha)\hat S(r,\theta)\rho_T\hat S^\dag(r,\theta)
\hat D^{\dag}(\alpha),\label{eqn:rho_param}
\end{eqnarray}
where 
\begin{eqnarray}
\rho_T=\frac{1}{{\bar n}+1}\sum_{n=0}^{\infty}\left(\frac{{\bar n}}{{\bar n}+1}
\right)^{\mkern-2mu n}|n\rangle\langle n|
\end{eqnarray}
is a thermal state of mean photon number ${\bar n}$, $\hat S\left(r,\theta\right)\equiv
\exp\mkern-2mu\left[(r/2)(e^{i\theta}\hat a^{\dag2}-e^{-i\theta}\hat a^2)\right]$ 
is the squeezing operator, and $\hat D(\alpha)\equiv\exp\mkern-2mu\left(\alpha
\hat a^\dag-\alpha^*\hat a\right)$ is the displacement operator. The covariance matrix
of the thermal state is diagonal---${\bm \Gamma}_T={\rm diag}\{\gamma,
\gamma\}$, $\gamma=2{\bar n}+1$---hence the covariance matrix of the state $\rho$
takes the factorized form [see Eq.~(\ref{eqn:cov_trans})]   
\begin{eqnarray}
{\bm \Gamma}={\bm \Phi}{\bm S}
\begin{pmatrix}
{\gamma}&{0}\cr
{0}&{\gamma}
\end{pmatrix}\mkern-3mu
{\bm S}^T{\bm \Phi}^T,
\end{eqnarray}
where
\begin{equation}
\Phi=
\begin{pmatrix}
{\cos\theta}&{-\sin\theta}\cr{\sin\theta}&{\cos\theta}
\end{pmatrix},
\qquad{\bm S}=
\begin{pmatrix}
{\sqrt{s}}&{0}\cr{0}&{1/\sqrt{s}}
\end{pmatrix},
\end{equation}
and the parameters $\theta$ and $s=e^{2r}$ denote the direction and degree of squeezing, 
respectively. The displacement operator leaves the covariance matrix unchanged, 
but fixes the mean field amplitude as $\langle a\rangle=\alpha=\alpha_x+i\alpha_y$. Thus,
the general single-mode Gaussian state is parameterized by the real numbers 
$\{\gamma,s,\theta,\alpha_x,\alpha_y\}$. The purity of the state is determined solely
by $\gamma$, with $\det{\bm\Gamma}=\gamma^2\geq~1$; a pure state corresponds
to the case $\gamma=1$. Without loss of generality, the squeezing parameter can be
restricted to $s\ge1$. A state is squeezed whenever $\gamma/s<1$. 

\subsection{Fidelity}
Given two quantum states, $\rho_1$ and $\rho_2$, the fidelity $F$ 
is defined by 
\begin{equation}
F(\rho_1,\rho_2)={\rm tr}\sqrt{\mkern-3mu\sqrt{\rho_1}\mkern3mu\rho_2\sqrt{\rho_1}}.
\end{equation} 
The fidelity $F$ is continuous with respect to $\rho_1$ and $\rho_2$, and concave, i.e.,
$F\left(\Sigma_ip_i\rho_i,\sigma\right)\ge\Sigma_ip_iF(\rho_i,\sigma)$ with $\Sigma_ip_i=1$.
It has the following properties as a measure of the closeness of the two states:
\par
\begin{itemize}
\item[(i)]
$F(\rho_1,\rho_2)=1$ if and only if $\rho_1=\rho_2$; more generally,
$0\le F(\rho_1,\rho_2)\le 1$. 
\item[(ii)]
$F(\rho_1,\rho_2)$ is symmetric, i.e., $F(\rho_1,\rho_2)=F(\rho_2,\rho_1)$.
\item[(iii)]
$F(\rho_1,\rho_2)=\sqrt{\langle\Psi_1|\rho_2|\Psi_1\rangle}$ 
when one of the states is pure---i.e., when $\rho_1=|\Psi_1\rangle\langle\Psi_1|$.
\item[(iv)]
$F(\rho_1,\rho_2)$ does not change under a unitary transformation $\hat U$---i.e., 
$F(\hat U\rho_1\hat U^\dag,\hat U\rho_2\hat U^\dag)=F(\rho_1,\rho_2)$.
\end{itemize}
Although the fidelity itself is not a metric, the angle between two states defined by 
$A(\rho_1,\rho_2)\equiv\cos^{-1}[F(\rho_1,\rho_2)]$ satisfies the triangular inequality 
$A(\rho,\sigma)\le A(\rho,\tau)+A(\tau,\sigma)$.

The main difficulty in calculating the fidelity comes from the square root of operators. 
For Gaussian states the difficulty is readily resolved, however, since the characteristic
function of the square-root of a Gaussian state is also Gaussian. Specifically, by a
successive use of the composition rule in the position representation, $\langle x
|\rho_1\rho_2|y\rangle=\int dz\langle x|\rho_1|z\rangle\langle z|\rho_2|y\rangle$,
with $\langle x|\rho|y\rangle=\exp[-(ax^2+dy^2+2bxy)+lx+ky+g]$, Scutaru \cite{Scutaru}
showed that the quantum fidelity of two Gaussian states, with covariance matrices
${\bm \Gamma}_i$ and mean amplitudes ${\bm \alpha}_i\equiv(\alpha_{ix},\alpha_{iy})$
($i=1,2$), is given by
 
\begin{eqnarray}
F=\sqrt{\frac{2}{\sqrt{\Delta+\delta}-\sqrt{\delta}}}\mkern3mu
\exp\mkern-2mu\left[-{\bm\beta}^T({\bm\Gamma}_1+{\bm\Gamma}_2)^{-1}{\bm\beta}\right],
\label{eqn:fidelity}
\end{eqnarray}
where 
\begin{eqnarray}
\Delta&=&\det({\bm\Gamma}_1+{\bm\Gamma}_2),\nonumber\\
\delta&=&(\det{\bm\Gamma}_1-1)(\det{\bm\Gamma}_2-1),\nonumber\\
{\bm \beta}&=&{\bm \alpha}_2-{\bm \alpha_1}.
\end{eqnarray}
It is also possible to obtain the fidelity by employing the Uhlmann theorem \cite{Uhlmann}, 
which states that
\begin{equation}
F(\rho_1,\rho_2)={\rm max}_{\{|\Psi\rangle,|\phi\rangle\}}|\langle\Psi|\phi\rangle|,
\end{equation} 
where the maximum is taken over all possible purifications, $|\Psi\rangle$ and
$|\phi\rangle$, of states $\rho_1$ and $\rho_2$, respectively, in an extended Hilbert
space $H_R\otimes H_S$---i.e., where $\rho_1={\rm tr}_R|\Psi\rangle\langle\Psi|$ and
$\rho_2={\rm tr}_R|\phi\rangle\langle\phi|$.

\section{Distinguishing Gaussian states by homodyne detection}
When homodyne measurement of the quadrature amplitude $\hat X_\phi=\hat X\cos\phi
+\hat Y\sin\phi$ is performed on a Gaussian state $\rho_i$ (parameterized by
$\{\gamma_i,s_i,\theta_i,\alpha_{ix},\alpha_{iy}\}$) the probability distribution
$p_i(x_\phi)$ is given by
\begin{eqnarray}
p_i(x_\phi)&=&\int dx_{\phi+\pi/2}W_i\left(x_\phi,x_{\phi+\pi/2}\right),
\label{eqn:prob_def}
\end{eqnarray}
where $W_i$ is the Wigner function
\begin{eqnarray}
W_i(\beta)=\frac{1}{\pi^2}\int d^2\lambda\mkern2mu C_i(\lambda)e^{\beta\lambda^*
-\beta^*\lambda},
\label{eqn:Wigner}
\end{eqnarray}
with $C_i$ the characteristic function; in this expression $\beta=\beta_x+i\beta_y$ is
a complex variable, related to $x_\phi$ via
\begin{eqnarray}
\begin{pmatrix}
{x_\phi}\cr{x_{\phi+\frac{\pi}{2}}}
\end{pmatrix}=
\begin{pmatrix}
{\cos\phi}&{\sin\phi}\cr{-\sin\phi}&{\cos\phi}
\end{pmatrix}\mkern-3mu
\begin{pmatrix}
{\beta_x}\cr{\beta_y}
\end{pmatrix}, 
\end{eqnarray}
and $\lambda=\lambda_x+i\lambda_y$ is also complex, with [equivalent to
Eq.~(\ref{eqn:characteristic})]
\begin{eqnarray} 
C_i(\lambda)\equiv{\rm tr}\{\rho_i\hat D(\lambda)\}=e^{\lambda\alpha_i^*-\lambda^*
\alpha_i}e^{-\frac{1}{2}{\bm \lambda}{\bm \Gamma}_i{\bm \lambda}^T},
\label{eqn:char_again}
\end{eqnarray} 
where ${\bm \lambda}$ denotes the row vector $(\lambda_y,-\lambda_x)$ and
$\hat D(\lambda)$ is the displacement operator. From
Eqs.~(\ref{eqn:prob_def})--(\ref{eqn:char_again}), we obtain the probability distribution 
\begin{eqnarray}
p_i(x_\phi)=\sqrt{\frac{2}{\pi B_i}}\exp\mkern-2mu\left[-\frac{2}{B_i}
(x_\phi-\alpha_{i,\phi})^2\right],
\label{eqn:prob_dist}
\end{eqnarray}
with
\begin{eqnarray}
B_i&=&\gamma_i[s_i\cos^2(\phi-\theta_i)+s_i^{-1}\sin^2(\phi-\theta_i)],\nonumber\\
\alpha_{i,\phi}&=&\alpha_{ix}\cos\phi+\alpha_{iy}\sin\phi.
\end{eqnarray}

Working from Eq.~(\ref{eqn:prob_dist}) we find that the overlap $I_\phi$ of the probability
distributions of two Gaussian states, $\rho_1$ and $\rho_2$, is given by
\begin{eqnarray}
I_\phi&\equiv&\int dx_\phi\sqrt{p_1(x_\phi)}\sqrt{p_2(x_\phi)}\nonumber\\
&=&\sqrt{\frac{2}{B_1+B_2}}\mkern2mu(B_1B_2)^{\frac{1}{4}}\exp\mkern-3mu
\left(-\frac{1}{B_1+B_2}\beta_\phi^2\right),
\label{eqn:overlapp}
\end{eqnarray}
where $\beta_\phi=\alpha_{2,\phi}-\alpha_{1,\phi}$. Restricting ourselves, then, to
states of the same mean ($\beta_\phi=0$), we arrive at the result
\begin{eqnarray}
I_\phi=f\left(\frac{B_2}{B_1}\right),\qquad f(x)\equiv\frac{\sqrt{2}x^{1/4}}{\sqrt{1+x}},
\label{eqn:overlap}
\end{eqnarray}
and the fidelity [Eq.~(\ref{eqn:fidelity})] is 
\begin{eqnarray}
F=\left(\frac{2}{\sqrt{\Delta+\delta}-\sqrt{\delta}}\right)^{\mkern-2mu1/2},
\label{eqn:ffidelity}
\end{eqnarray}
where, in terms of the parameters $\{\gamma_i,s_i,\theta_i\}$ ($i=1,2$),
\begin{eqnarray}
\Delta&=&\gamma_1^2+\gamma_2^2+{\textstyle\frac{1}{2}\displaystyle}\gamma_1\gamma_2
D(s_1,s_2,\tilde\theta),\nonumber\\
\delta&=&\left(\gamma_1^2-1\right)\left(\gamma_2^2-1\right),
\label{eqn:coeff}
\end{eqnarray}
with $\tilde{\theta}\equiv\theta_2-\theta_1$ and
\begin{equation}
D(s_1,s_2,\tilde\theta)\equiv s_{1+}s_{2+}-s_{1-}s_{2-}\cos2\tilde\theta,
\label{eqn:D_param}
\end{equation}
where
\begin{equation}
s_{i\pm}\equiv s_i\pm s_i^{-1}\qquad(i=1,2).
\end{equation}
Note that the function $f(x)$ [Eq.~(\ref{eqn:overlap})] is concave. Therefore in any
interval $x\in[a,b]$, the minimal value of $f(x)$ is realized at one of the endpoints,
$x=a$ or $x=b$. Our task is to compare the minimal value of $I_\phi$ with the
fidelity~(\ref{eqn:ffidelity}) to determine under what conditions they are
equal. To illustrate how the minimum is achieved, we first consider some simple cases.

\subsection{$s_1=1$ and $s_2\ge1$} 
In this case the Wigner distribution of the state $\rho_1$ is a radially symmetric
Gaussian [Fig.~1 (a)]. It is easy to see, then, that the minimum $I_\phi$ is achieved
for a homodyne measurement in either the ``unsqueezed'' or ``squeezed'' direction of the
state $\rho_2$---i.e., for $\phi=\theta_2$ or $\phi=\theta_2+\pi/2$ (note that squeezing in
the strict sense requires $\gamma_2/s_2<1$). The minimum is therefore achieved with
either $B_2/B_1=(\gamma_2s_2)/\gamma_1$ or $B_2/B_1=\gamma_2/(\gamma_1s_2)$. 

(a) $\gamma_1\ge\gamma_2$:
In this case the minimum occurs for $\phi=\theta_2+\pi/2$
and $B_2/B_1=\gamma_2/(\gamma_1s_2)$. Then, by comparing $f(B_2/B_1)$ with $F$, in a
tedious but straightforward manner, it is found that the minimum is equal to $F$
when (i) $\gamma_1=\gamma_2=1$ (in which case both states are pure) or (ii) $s_2=
(\gamma_1-\gamma_1^{-1})/(\gamma_2-\gamma_2^{-1})$, with $\gamma_1,\gamma_2>1$ (both
states are mixed). The equality does not hold and homodyne detection is not
optimal when one state is pure and the other is mixed ($\gamma_1=1<\gamma_2$). 
If $\gamma_1=\gamma_2>1$, then $s_2$ must be unity according to (ii), and
the two states are identical.

(b) $\gamma_1\le\gamma_2$: 
In this case the minimum occurs for $\phi=~\theta_2$ and $B_2/B_1=(\gamma_2s_2)/\gamma_1$.
Checking for equality with $F$, in a similar fashion, we obtain the
conditions (i) $\gamma_1=\gamma_2=1$ and (ii)~$s_2=(\gamma_2-\gamma_2^{-1})/
(\gamma_1-\gamma_1^{-1})$, with $\gamma_1,\gamma_2>1$.

In summary, homodyne detection is found to be optimal for distinguishing two
Gaussian states when 
\begin{eqnarray}
&&\gamma_1=\gamma_2=1,\qquad\qquad\qquad\quad\mkern6mu\hbox{(i)}\nonumber\\
&&s_{2+}=\Gamma_{12}\quad(\gamma_1,\gamma_2>1),\qquad\hbox{(ii)}
\label{eqn:condition1}
\end{eqnarray}
where
\begin{eqnarray}
\Gamma_{12}\equiv\frac{\gamma_2-\gamma_2^{-1}}{\gamma_1-\gamma_1^{-1}}
+\frac{\gamma_1-\gamma_1^{-1}}{\gamma_2-\gamma_2^{-1}}.
\end{eqnarray}
Note that when two pure states are compared [case (i)], the equality $I_\phi=F$ holds
for both phase angles, $\phi=\theta_2$ and $\phi=\theta_2+\pi/2$. 

\begin{figure}
\includegraphics*[width=2.25in,keepaspectratio=true]{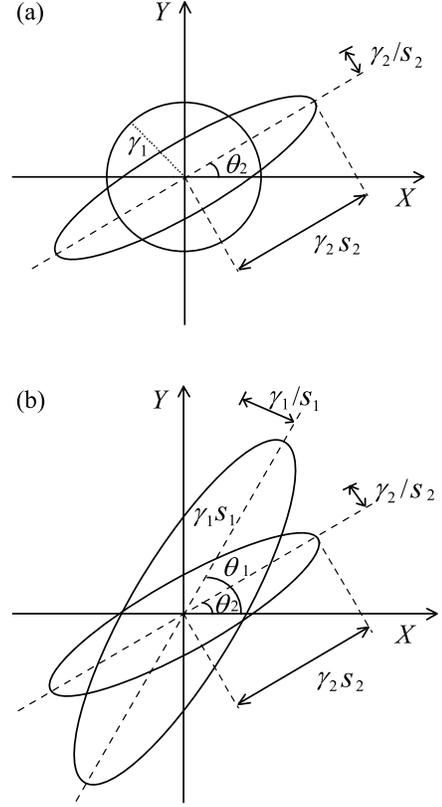}
\caption{Parameterization for two Gaussian states with the same mean. Contours of
the Wigner function are plotted: (a)~the special case where state $\rho_1$ is radially
symmetric in phase space ($s_1=1$), (b) the general case.}
\label{fig:fig1}
\end{figure}

\subsection{$s_1\ge1$ and $s_2\ge1$}
In this more general case [Fig.~1 (b)] it is not straightforward to see which phase
angle gives the minimal value of $I_\phi$. Nevertheless, we can obtain the conditions
under which the equality ${\rm min}_{\{\phi\}} I_\phi=F$ holds in the following way.

The question is whether there exists an angle $\phi$ for which $I_\phi=F$. Note first,
from Eq.~(\ref{eqn:overlap}), that the equality is equivalently
$B_2/B_1=[F^{-2}\pm(F^{-4}-1)^{1/2}]^2$, an equation that can be arranged in the form 
\begin{eqnarray}
A_1\sin2\phi+A_2\cos2\phi+A_3=0, 
\label{eqn:sinusoidal}
\end{eqnarray} 
by using 
\begin{eqnarray}
\frac{B_2}{B_1}=\frac{\gamma_2\left(s_{2+}+s_{2-}\cos2(\phi-\theta_2)\right)}
{\gamma_1\left(s_{1+}+s_{1-}\cos2(\phi-\theta_1)\right)}. 
\end{eqnarray}
Equation~(\ref{eqn:sinusoidal}) has a solution if and only if the inequality
$A_1^2+A_2^2-A_3^2\ge0$ is satisfied. In our case, this condition may be expressed as 
\begin{eqnarray}
2\Upsilon^2-\Upsilon D(s_1,s_2,\tilde{\theta})+2\le0,
\label{eqn:inequality}
\end{eqnarray}
where
\begin{eqnarray}
\Upsilon=(\gamma_1/\gamma_2)[F^{-2}\pm(F^{-4}-1)^{1/2}]^2.
\end{eqnarray}
Importantly, inequality~(\ref{eqn:inequality}) has only {\it three} independent
parameters, $\gamma_1$, $\gamma_2$, and $D(s_1,s_2,\tilde{\theta})$.
Thus, the condition for ${\rm min}_{\{\phi\}} I_\phi=F$ may be expressed 
in terms of $\gamma_1$ ,$\gamma_2$, and $D(s_1,s_2,\tilde{\theta})$; in other
words, we need not treat the parameters $s_1$, $s_2$ and $\tilde{\theta}$
independently. We can deduce general conditions by considering a few simple cases,
such as $\tilde{\theta}=0$ and $\tilde\theta=\pi/2$, instead of solving the complicated
inequality~(\ref{eqn:inequality}). 

Let us first consider the case $\tilde{\theta}=\theta_2-\theta_1=0$---i.e., the two
states are squeezed in the same direction. Since the extremal values of $B_2/B_1$
are then $\gamma_2s_2/\gamma_1s_1$ and $\gamma_2s_1/\gamma_1s_2$, this case is similar
to that in the previous subsection.  Only condition (ii) of Sec.~IIIA is changed, 
with $s_2$ replaced by $s_2/s_1$ in Eq.~(\ref{eqn:condition1}), to obtain
\begin{eqnarray}
s_2s_1^{-1}+s_1s_2^{-1}=\Gamma_{12}.
\label{eqn:condition2}
\end{eqnarray}
Secondly, with $\tilde{\theta}=\pi/2$, we similarly have $\gamma_2/\gamma_1s_1s_2\le B_2/
B_1\le\gamma_2s_1s_2/\gamma_1$. Thus, condition (ii) in Sec.~IIIA is changed to 
\begin{eqnarray}
s_1s_2+(s_1s_2)^{-1}=\Gamma_{12}.
\label{eqn:condition3}
\end{eqnarray}
Note that, according to the definition~(\ref{eqn:D_param}), the term $s_2s_1^{-1}
+s_1s_2^{-1}$ in Eq.~(\ref{eqn:condition2}) is equal to $\frac{1}{2}D(s_1,s_2,0)$,
while the term $s_1s_2+(s_1s_2)^{-1}$ in Eq.~(\ref{eqn:condition3}) is equal to
$\frac{1}{2}D(s_1,s_2,\pi/2)$. Thus, we deduce the general condition ensuring
${\rm min}_{\{\phi\}}I_\phi=F$ in the following way. 

\begin{figure}
\includegraphics*[width=2.25in,keepaspectratio=true]{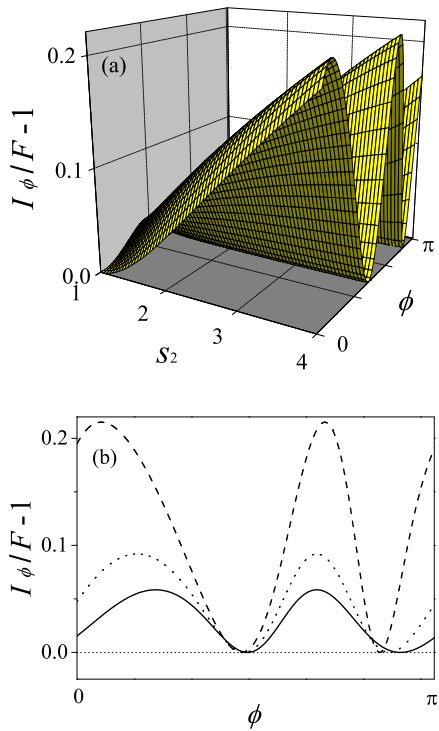}
\caption{Normalized difference of the probability overlap $I_\phi$ measured in homodyne
detection and the fidelity $F$ in the case of {\it pure} states ($\gamma_1=\gamma_2=1$), 
for $s_1=2$ and $\tilde{\theta}=\pi/3$: (a)~as a function of $s_2$ and $\phi$, 
(b) as a function of $\phi$ for $s_2=1.5$ (solid), $s_2=2$ (dotted), and $s_2=4$ (dashed). 
It is seen that there always exists an angle $\phi$ for which $I_\phi=F$. }
\label{fig:fig2}
\end{figure}

(i) $\gamma_1=\gamma_2=1$:
When one of the states is pure the other must be pure also in order to obtain equality.
Then, homodyne detection provides optimal distinguishability without regard to the parameters 
$\{s_1,s_2,\theta_1,\theta_2\}$. This case is demonstrated in Fig.~2. Thus, {\it two pure
Gaussian states are always optimally distinguished by homodyne detection\/}. Also, when one
state is pure and the other not ($\gamma_1=1<\gamma_2=1$) equality between
${\rm min}_{\{\phi\}}I_\phi$ and $F$ never holds, as illustrated by Fig.~3: {\it one pure
and one mixed Gaussian state are never optimally distinguished by homodyne detection\/}.

(ii) $\gamma_1>1$ and $\gamma_2>1$: In the case of two mixed states, the condition 
\begin{eqnarray}
D(s_1,s_2,\tilde{\theta})=2\Gamma_{12}
\label{eqn:condition4}
\end{eqnarray}
must be satisfied \cite{nha}. The case is illustrated by Fig.~4. It turns out that
condition~(\ref{eqn:condition4}) is very restrictive and the parameters satisfying it
occupy a volume of measure zero in the 5-dimensional parameter space
$\{\gamma_1,\gamma_2,s_1,s_2,\tilde{\theta}\}$. 

\begin{figure}
\includegraphics*[width=2.25in,keepaspectratio=true]{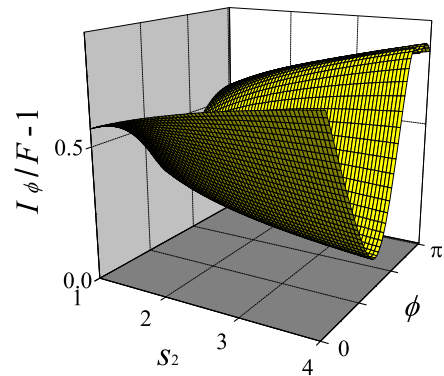}
\caption{Normalized difference of the probability overlap $I_\phi$ measured in homodyne detection 
and the fidelity $F$ when one state is {\it pure} and the other is {\it mixed} ($\gamma_1=1$,
$\gamma_2=4$), for $s_1=2$ and $\tilde{\theta}=\pi/3$. The plot shows that the equality $I_\phi=F$
does not hold for any $s_2$.}
\label{fig:fig3}
\end{figure}

\subsection{Gaussian states of different mean}
Finally, we comment on the case where the means of the two states are not
the same. In this situation, the measurement direction $\phi$ yielding the minimal value of $I_\phi$
is a function of the five parameters $\{s_1,s_2,\theta_2-\theta_1,{\bm\alpha}_2-{\bm\alpha}_1\}$. 
If we confine ourselves to pairs of Gaussian states that are radially symmetric in phase space
($s_1=s_2=1$), then it is easily shown that homodyne detection is optimal if and only if
$\gamma_1=\gamma_2$; the variances of the two states must be the same. Examples include all pairs
of coherent states. 

\begin{figure}
\includegraphics*[width=2.25in,keepaspectratio=true]{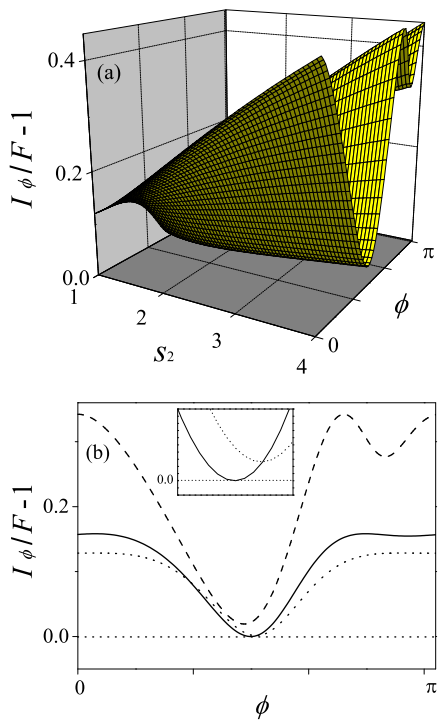}
\caption{Normalized difference of the probability overlap $I_\phi$ measured in homodyne detection 
and the fidelity $F$ in the case of two {\it mixed} states ($\gamma_1=2$, $\gamma_2=4$), for
$s_1=2$ and $\tilde{\theta}=\pi/3$: (a)~as a function of $s_2$ and $\phi$, (b) as a
function of $\phi$ for $s_2=1.4$ (solid), $s_2=1.1$ (dotted), and $s_2=3$ (dashed). 
It is seen that there exists an angle $\phi$ for which $I_\phi=F$ only when
Eq.~(\ref{eqn:condition4}) is satisfied, i.e., for $s_2=1.4$. The inset expands the
region around the minimum.}
\label{fig:fig4}
\end{figure}

\section{Summary and Discussion} 
We have investigated the conditions under which homodyne detection can
optimally distinguish, in an information-theoretic sense, two Gaussian states
of the same mean. We found that two pure Gaussian states are always optimally
distinguished by homodyne detection. On the other hand, if one of the states is
mixed, homodyne detection is optimal only for pairs of states satisfying
Eq.~(\ref{eqn:condition4}), a condition satisfied within a region of measure
zero in the parameter space. 

In general, when one of the states to be compared is pure, say $\rho_1=|\Psi_1\rangle
\langle\Psi_1|$, an optimal POVM is provided by the two-component measurement
$E_0=|\Psi_1\rangle\langle\Psi_1|$, $E_1=I-|\Psi_1\rangle\langle\Psi_1|$, for which
it is easy to see that $\sum_{m=0}^1\sqrt{p_1(m)}\sqrt{p_2(m)}=\sqrt{\langle\Psi_1|
\rho_2|\Psi_1\rangle}=F$. For example, when $\rho_1$ is the vacuum state, the measurement
can be implemented with a photodetector, by discriminating between no-click ($E_0$)
and click ($E_1$) events. Such a POVM is clearly state-dependent, however,
and requires prior knowledge of the states involved. It is not, therefore, of general
practical use. (Reference~\cite{Fuchs} also constructs an optimal POVM that
depends on the states to be distinguished.) Our approach, on the other hand, considers
a predetermined class of measurements, namely, homodyne measurements, and we show
that the considered measurements can always optimally distinguish two {\it pure}
Gaussian states; the prior knowledge is only that the two states are pure. 
Of course, our approach is only possible because the optimal POVM is not, in general,
unique. In spite of its merit, the practical applicability of our
result is still rather restricted. For example, setting aside technical
imperfections in homodyne measurement, both imperfect quantum teleportation and quantum
cloning output a mixed state, even when the input state is pure.   

We have shown that two pure Gaussian
states of the same mean can be optimally distinguished by a pure CV measurement\cite{Nha1}
but we considered a restricted class of homodyne detection only, and the question remains open as to
whether or not it is possible to define a generalized CV measurement to distinguish
between mixed states. 
One possible generalization is the POVM measurement where each component is given by 
$E_\alpha=\frac{1}{\pi}U^{\dag}|\alpha\rangle\langle\alpha|U$ and 
$\int d^2\alpha E_\alpha =I$. 
Here, $U$ is an arbitrary unitary operator and $|\alpha\rangle$ is a coherent state. 
This measurement can be implemented as follows. 
First, the signal field $\rho_s$ is subject to the unitary evolution $U$ and 
then mixed at a 50:50 beam splitter with an auxiliary mode in the vacuum state $|0\rangle$. 
Balanced homodyne detections are performed on the two output fields from the beam splitter, 
one for the $\hat X_{0}$ quadrature and the other for the $\hat X_{\pi/2}$ quadrature 
(or, more generally, one for $\hat X_{\phi}$ and the other for $\hat X_{\phi+\pi/2}$).
The joint probability then corresponds to 
${\rm tr \{E_\alpha \rho_s}\}=
\frac{1}{\pi}\langle\alpha|U\rho_sU^{\dag}|\alpha\rangle$. 

Now, if the unitary operation $U$ is restricted to the symplectic transformations, 
the above POVM measurements correspond to projections into the pure Gaussian states
\cite{Eisert}. 
One can consider, for example, the squeezing operations $U=S(r,\theta)$, for which, 
in the limit $r\rightarrow\infty$, the measurement becomes that of homodyne detection 
of the quadrature amplitude $\hat X_{\theta}$, 
the measurement considered throughout this paper. 
On the other hand, in the opposite limit, $r\rightarrow0$, it becomes the so called 
heterodyne detection where the $Q$-distribution of the state is measured~\cite{Leonhardt}. 
In this generalization, it would be interesting to prove or disprove
the conjecture that the measurement scheme with $r\rightarrow\infty$ is optimal 
among all $S(r,\theta)$-parameterized POVMs. 
Such considerations appear of value to applications
beyond the present limited context of the distinguishability of states and 
they are thus left for future work.

This work was supported by the NSF under Grant No.\ PHY-0099576 and
by the Marsden Fund of the RSNZ.


\begin{references}
\bibitem{nha} Condition~(\ref{eqn:condition4}) corresponds to equality in
Eq.~(\ref{eqn:inequality}), and one might wonder if there can be a range of parameters
in the case of strict inequality. It is readily shown, however, that there cannot.
\end{references}
\end{document}